\documentclass[showpacs,aps,prd,nofootinbib,floatfix,amsmath,amssymb]{revtex4}
\usepackage{graphicx}
\usepackage{graphics}
\usepackage{epsfig}
\usepackage{axodraw}
\newcommand{\bea}{\begin{eqnarray}}
\newcommand{\eea}{\end{eqnarray}}
\begin{document}

\makeatletter
\newbox\slashbox \setbox\slashbox=\hbox{$/$}
\newbox\Slashbox \setbox\Slashbox=\hbox{\large$/$}
\def\pFMslash#1{\setbox\@tempboxa=\hbox{$#1$}
  \@tempdima=0.5\wd\slashbox \advance\@tempdima 0.5\wd\@tempboxa
  \copy\slashbox \kern-\@tempdima \box\@tempboxa}
\def\pFMSlash#1{\setbox\@tempboxa=\hbox{$#1$}
  \@tempdima=0.5\wd\Slashbox \advance\@tempdima 0.5\wd\@tempboxa
  \copy\Slashbox \kern-\@tempdima \box\@tempboxa}
\def\FMslash{\protect\pFMslash}
\def\FMSlash{\protect\pFMSlash}
\def\miss#1{\ifmmode{/\mkern-11mu #1}\else{${/\mkern-11mu #1}$}\fi}
\makeatother

\title{The anapole moment in scalar quantum electrodynamics}
\author{A. Bashir$^{(a)}$, Y. Concha--S\' anchez$^{(a,b)}$, M.E. Tejeda-Yeomans
$^{(c)}$, J. J. Toscano$^{(a,b)}$}
\address{$^{(a)}$Instituto de F\'{\i}sica y Matem\' aticas, Universidad
Michoacana de San Nicol\' as de Hidalgo, Edificio C-3, Ciudad
Universitaria, C.P. 58040, Morelia, Michoac\' an, M\' exico.\\
$^{(b)}$Facultad de Ciencias F\'{\i}sico Matem\' aticas,
Benem\'erita Universidad Aut\'onoma de Puebla, Apartado Postal
1152, Puebla, Puebla, M\'exico.\\
$^{(c)}$Departamento de F\'isica, Universidad de Sonora, Boulevard Luis
Encinas J. y Rosales, Colonia Centro, C.P. 83000, Hermosillo, Sonora,
M\'exico.
}

\begin{abstract}
The anapole moment of a charged scalar particle is studied
in a model independent fashion, using the effective Lagrangian technique,
as well as radiatively within the context of scalar quantum
electrodynamics (SQED). It is shown that this gauge structure is characterized
by a non renormalizable interaction, which is radiatively generated at
the one--loop. It is found that the resulting anapole moment for
off-shell particles,
though free of ultraviolet divergences, is gauge
dependent and thus it is not a physical observable. We also study some of its kinematical limits. In particular, it is shown that its value comes out to be zero when all particles are on--shell.
\end{abstract}


\maketitle In a previous communication by some of us, a one--loop SQED
calculation for the 3--point off--shell $\phi^-\phi^+\gamma$ Green
function in arbitrary gauge and dimension was reported~\cite{BCD}.
It was shown that a new independent gauge structure arises at this
level. In this brief report, we show that this non renormalizable
gauge structure corresponds to the anapole moment of a charged scalar
particle. To the best of our knowledge, this study has not been carried
out in the literature so far. The anapole moment, first
introduced by Zeldovich~\cite{Zeldovich}, is an electromagnetic
property that vanishes for on--shell photons and always arises as
a quantum fluctuation at one--loop or higher orders
within the context of renormalizable theories. It is
the only electromagnetic property of  Majorana
particles.
The anapole moment is best known from the studies of Majorana
neutrinos~\cite{MN} and the standard model (SM) $Z$ boson~\cite{ZB},
although it is a characteristic of any type of
particle\footnote{The only exception is the neutral scalar
particle that coincides with its own antiparticle, which has no
electromagnetic properties.}. Its systematic
study for Majorana particles
including fields of higher spin has been carried out in~\cite{FB}.
The anapole moment of spin $1/2$ non Majorana particles
has also been the subject of interest in the literature~\cite{Musolf}.
In the present work, we focus on the anapole moment associated with a
scalar particle within the context of SQED. We study this
electromagnetic property of charged scalars in two different ways.
Firstly, using the effective Lagrangian formalism~\cite{EL}, we
demonstrate the existence of this gauge structure in a model--independent way.
Secondly, using general results reported in~\cite{BCD} for a
charged scalar, we show that this off--shell electromagnetic
structure arises at the one--loop level. Apart from its intrinsic theoretical
interest, there are phenomenological motivations to study the electromagnetic
properties of scalar particles, as it is expected that new physics show up during the experiments that will be carried out at the Large Hadron Collider (LHC).
In particular, if there is a Higgs boson, it is almost certain to be found at
this collider and its mass measured by the ATLAS~\cite{ATLAS} and
CMS~\cite{CMS} experiments. If discovered, answers to many related questions
would have to be sought. For instance, does it differ from the one predicted
by the Standard Model (SM)? Are there more than one Higgs particles?
Therefore, it is important to study all types of physical properties of scalar
particles, including those that arise as a quantum fluctuation, as is
the case of the anapole moment.

As it is well known, the renormalizable structure of SQED induces
three-- and four--point $\phi^-\phi^+\gamma$ and $\phi^-\phi^+\gamma \gamma$
couplings at the level of the classical action.
However, it is not difficult to convince ourselves that these
couplings develop new components of non renormalizable type at the
level of the quantum action. The corresponding effective theory
incorporates $U_e(1)$--invariant terms of up to dimension six.
To that order, one can write an effective Lagrangian of
the form
\begin{equation}
{\cal L}_{eff}={\cal L}_{SQED}+\frac{\alpha_{D}}{\Lambda^2}{\cal
O}_{D}+\frac{\alpha_{DD}}{\Lambda^2}{\cal O}_{DD},
\end{equation}
where $\Lambda$ is some relevant energy scale and ${\cal L}_{SQED}$ represents
the renormalizable Lagrangian given by~:
\begin{equation}
{\cal L}_{SQED}=(D_\mu \phi^+)^\dag (D^\mu
\phi^+)-V(\phi^-,\phi^+)-\frac{1}{4}F_{\mu \nu}F^{\mu \nu}+{\cal
L}_{GF} \;.
\end{equation}
In these expressions,
$D_\mu=\partial_\mu-ieA_\mu$ is the electromagnetic covariant
derivative, $V(\phi^-,\phi^+)$ is the scalar potential, ${\cal
L}_{GF}$ is the gauge--fixing term, and the ${\cal O}_{i}$ are
$U_e(1)$--invariant structures of dimension six given
by~\footnote{The $U_e(1)$--invariant $(D_\mu \phi^+)(D_\nu
\phi^+)^\dag F^{\mu \nu}$ term is not independent, as it is
related to ${\cal O}_{D}$ and ${\cal O}_{DD}$ through a surface
term.}
\begin{eqnarray}
{\cal O}_{D}&=&i\Big[\phi^-D_\nu \phi^+-\phi^+(D_\nu \phi^+)^\dag
\Big] \; \partial_\mu F^{\mu \nu}, \\
{\cal O}_{DD}&=&i\Big[\phi^-(D_\mu D_\nu \phi^+)-\phi^+(D_\mu
D_\nu \phi^+)^\dag \Big] \; F^{\mu \nu}.
\end{eqnarray}
As is evident, both ${\cal O}_{D}$ and ${\cal O}_{DD}$
contribute to the 4--point vertex, but only ${\cal O}_{D}$
contributes to the 3--point coupling. Thus, the most general
electromagnetic gauge structure for the $\phi^-\phi^+\gamma$
vertex is given by
\begin{equation}
{\cal L}_{\phi^-\phi^+\gamma}=ieJ_\mu A^\mu
+\frac{i\alpha_D}{\Lambda^2}J_\mu \partial_\nu F^{\nu \mu},
\end{equation}
where $J_\mu$ is the electromagnetic current given by:
\begin{equation}
J_\mu=\phi^-\partial_\mu \phi^+-\phi^+\partial_\mu \phi^-.
\end{equation}
From the above Lagrangian, one can construct the most general
3--point vertex function, which can be written as:
\begin{equation}
{\Gamma_\mu}=f_Q(q^2,p^2,k^2)(k+p)_\mu+\frac{f_A(q^2,p^2,k^2)}{\Lambda^2}
\Big[q^2(k+p)_\mu-(k^2-p^2)q_\mu \Big],
\end{equation}
where $f_Q(q^2,p^2,k^2)$ and $f_A(q^2,p^2,k^2)$ are the form
factors associated with the monopole and anapole moments of the
$\phi^\pm$ scalar boson, respectively. In the context of the
renormalizable theory, $f_Q(q^2,p^2,k^2)=1$ and
$f_A(q^2,p^2,k^2)=0$ at the level of the classical action. As we will
show below, the anapole moment is generated at the one--loop level
by the renormalizable theory. The new gauge structure satisfies
$  q^\mu  [q^2(k+p)_\mu-(k^2-p^2)q_\mu ]=0$ ensuring the
Ward identity $
\label{wi} q^\mu \Gamma_\mu =S^{-1}(k^2)-S^{-1}(p^2)$ is conserved,
$S^{-1}$ being the scalar propagator. To be able to make quantitative
statements about the anapole form factor, we need to know
$f_A(q^2,p^2,k^2)$ explicitly.

\vspace{0.5cm}
\begin{center}
\begin{picture}(2000,70)(13,-30)
\SetScale{0.78} \SetWidth{1.2}
\CCirc(90,15){5}{Black}{Black}
\SetColor{Red} \Photon(25,15)(85,15){4}{7}
\Photon(166,15)(226,15){4}{7} \SetColor{Black}
\LongArrow(65,25)(45,25) \PText(55,5)(0)[]{q=k - p}
\LongArrow(206,25)(186,25) \PText(196,5)(0)[]{q=k - p}
\LongArrow(326,25)(306,25) \PText(324,5)(0)[]{q=k - p}
\LongArrow(386,5)(386,25) \PText(396,15)(0)[]{w}
\Text(45,30)[]{$A_{\mu}(q)$} \Text(90,50)[]{$\phi^+(p)$}
\Text(90,-30)[]{$\phi^-(k)$} \LongArrow(350,25)(360,35)
\LongArrow(360,-5)(350,5) \SetColor{Blue}
\ArrowLine(93,20)(143,65) \ArrowLine(143,-35)(93,10)
\ArrowLine(226,15)(276,65) \ArrowLine(276,-35)(226,15)
\SetColor{Black} \PText(146,75)(0)[]{p} \PText(146,-45)(0)[]{k}
\PText(146,15)(0)[]{ = } \PText(276,75)(0)[]{p}
\PText(276,-45)(0)[]{k} \PText(276,15)(0)[]{+} \SetColor{Red}
\Photon(296,15)(346,15){4}{7} \SetColor{Blue}
\ArrowLine(346,15)(396,65) \ArrowLine(396,-35)(346,15)
\SetColor{Black} \PText(396,75)(0)[]{p} \PText(396,-45)(0)[]{k}
\SetColor{Red}
\Photon(376,45)(376,-15){3}{6.5} \SetColor{Black}
\PText(416,15)(0)[]{+} \PText(350,40)(0)[]{p-w}
\PText(350,-10)(0)[]{k - w}
\LongArrow(466,25)(446,25) \LongArrow(586,25)(566,25)
\LongArrow(495,32)(505,42) \LongArrow(620,-10)(610,0)
\LongArrowArc(503,30)(10,257,-15) \LongArrowArc(626,0)(10,-13,95)
\PText(464,5)(0)[]{q=k - p} \PText(584,5)(0)[]{q=k - p}
\SetColor{Red} \Photon(426,15)(486,15){4}{7} \SetColor{Blue}
\ArrowLine(486,15)(536,65) \ArrowLine(536,-35)(486,15)
\SetColor{Black} \PText(536,75)(0)[]{p} \PText(536,-45)(0)[]{k}
\SetColor{Red} \PhotonArc(500,30)(20,225,45){2}{6.5}
\SetColor{Black} \PText(525,15)(0)[]{w} \PText(546,15)(0)[]{+}
\PText(500,50)(0)[]{p-w} \SetColor{Red}
\Photon(556,15)(606,15){3}{7} \SetColor{Blue}
\ArrowLine(606,15)(656,65) \ArrowLine(656,-35)(606,15)
\SetColor{Black} \PText(656,75)(0)[]{p} \PText(656,-45)(0)[]{k}
\SetColor{Red} \PhotonArc(623,0)(20,-45,135){2}{6.5}
\SetColor{Black} \PText(648,15)(0)[]{w}
\PText(613,-15)(0)[]{k-w} \SetScale{0.9} \SetColor{Black}
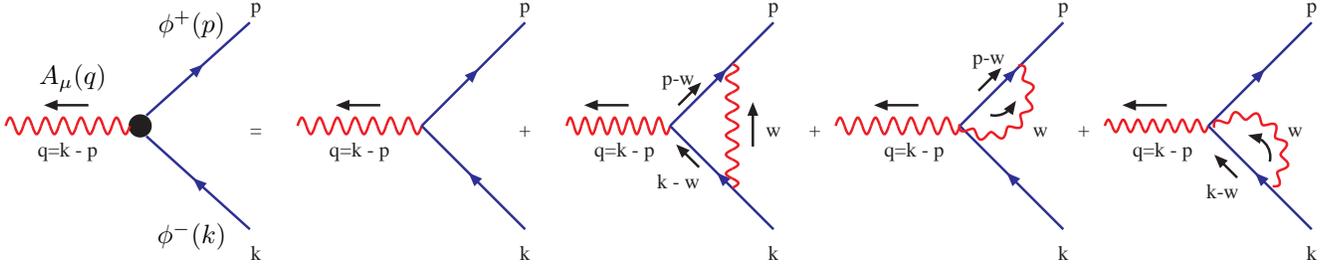
\Text(290,-70)[]{Figure 1: Diagrams contributing to the $\phi^+
\phi^- \gamma$ vertex up to one--loop order.}
\end{picture}
\end{center}
\vspace{2.0cm}

While the complete non perturbative
expression for the anapole moment is not known~\footnote{Transverse Ward
identities in SQED
fail to yield any information on $f_A(q^2,p^2,k^2)$.}, it is induced at
one--loop level. The one--loop contribution to the
$\phi^-\phi^+\gamma$ Green function is given through the diagrams
shown in Fig.~1. As it was shown on general grounds in~\cite{BCD},
the complete Green function $\Gamma_\mu$ is made of two independent
components, one proportional to the monopole moment $(k+p)_\mu$
and other associated with a part that satisfies the $q^\mu
\Gamma_\mu^T=0$ transverse condition. In the light of the
above discussion, this loop induced gauge
structure can be identified with the anapole moment~:
\bea
    \Gamma_\mu^T &=& \frac{f_A(q^2,p^2,k^2)}{\Lambda^2}
\Big[q^2(k+p)_\mu-(k^2-p^2)q_\mu \Big] \equiv \tau(k^2,p^2,q^2) T_{\mu}(k,p) \;,
\eea
where
\bea
T_{\mu}(k,p) &=& q^2(k+p)_\mu-(k^2-p^2)q_\mu \;.
\eea
The explicit general expression for $\tau(k^2,p^2,q^2)$ for the off--shell
particles in
arbitrary dimensions is given in~\cite{BCD}. In 4--dimensions, it is
\bea
\tau(k^2,p^2,q^2) &=& \frac{\alpha}{8 \pi \Delta^2}
\left[(k^2+p^2-4 k \cdot p) \left\{  k \cdot p J_0 + {\rm ln} \left[
\frac{q^4}{k^2 p^2}  \right]  \right\}  +
\frac{(k^2+p^2)q^2- 8 k^2 p^2}{p^2-k^2}
 \; {\rm ln} \left( \frac{k^2}{p^2} \right)
\right]   \nonumber  \\
&-& \frac{\alpha \xi^{\prime}}{8 \pi \Delta^2} \left[ k^2 p^2 J_0 +
\frac{2 k^2 p^2}{k^2-p^2} \; {\rm ln} \left( \frac{p^2}{k^2} \right) +
\frac{2 k \cdot p}{k^2 - p^2} \left\{ p^2  \; {\rm ln} \left( \frac{p^2}{q^2} \right)
+ k^2  \; {\rm ln} \left( \frac{q^2}{k^2} \right)
\right\}   \right] .
\eea
It is gauge--dependent but free of ultraviolet divergences. All such divergences reside in the
longitudinal part of the vertex. Here, we shall analyze different kinematic limits
of the transverse vertex in an arbitrary covariant gauge. We summarize our results as follows~:
\begin{itemize}

\item In the limit $q^2 \rightarrow 0$ for $k^2 \neq p^2$,
\bea
  \Gamma_{\mu}^T(q^2 \rightarrow 0) &= & -\frac{\alpha}{4 \pi} \, \frac{q_{\mu}}{(k^2-p^2)^2} \; {\rm Ln} \left( \frac{q^2}{p^2} \right) \; \times \nonumber \\
&&
\left[ 2(k^4-p^4)  + (k^2+p^2)^2  {\rm Ln} \left( \frac{p^2}{k^2} \right)
+ \xi^{\prime} \left\{  (k^4-p^4)  + 2 k^2 p^2  {\rm Ln} \left( \frac{p^2}{k^2} \right)       \right\}  \right]  \;,
\eea
where $\alpha=e^2/(4 \pi)$ and $\xi^{\prime}=1-\xi$. $\xi=0$ corresponds to the
Landau gauge and $\xi=1$ to the Feynman gauge. Note that this limit is logarithmically divergent and gauge dependent.

\item In the limit $k^2 \rightarrow p^2$ for $q^2 \neq 0$
\bea
   \Gamma_{\mu}^T(k^2 \rightarrow p^2) &=& -\frac{\alpha}{4 \pi} \, \frac{(k+p)_{\mu}}{(q^2 - 4 p^2)} \times \nonumber \\
&&
\hspace{-34mm}
\left[ (q^2-p^2)  \left( (2 p^2 - q^2) J_0 + 4 {\rm Ln}  \left( \frac{q^2}{p^2} \right)
   \right)
+ 2 (4 p^2 - q^2) + \xi^{\prime} \left\{-p^4 J_0 + (4 p^2 - q^2) - (2 p^2-q^2) \, {\rm Ln}  \left( \frac{q^2}{p^2} \right)
   \right\} \right] \; ,
\eea
where
\bea
  J_0 (k^2 \rightarrow p^2) &=& \frac{2}{\Delta} \left[ S_p\left(\frac{q^2 + 2 \Delta}{2 p^2} \right) -
 S_p\left(\frac{q^2 - 2 \Delta}{2 p^2} \right) + \frac{1}{2} \; {\rm Ln} \left( \frac{2 p^2 - q^2 - 2 \Delta}{2 p^2 - q^2 +
2 \Delta} \right) \;  {\rm Ln} \left( \frac{q^2}{p^2} \right)
   \right]  \;,
\eea
with $\Delta^2= q^2 (q^2 - 4 p^2)/4$ and
\bea
    S_p(x) &=& - \int_0^x dy \frac{{\rm Ln}(1-y)}{y} \;. \nonumber
\eea
Therefore, this limit, though gauge dependent, is perfectly finite, corresponding to the fact that
 there are
no kinematic singularities for the 3-point functions in scalar QED. The same is true for spinor
QED and QCD for  $k^2 \rightarrow p^2$.

\item Moreover, one can show that
\bea
 \Gamma_{\mu}^T  (k^2 \rightarrow p^2, q^2 \rightarrow 0) & = &
 \Gamma_{\mu}^T  (q^2 \rightarrow 0,   k^2 \rightarrow p^2) = 0  \;.
\eea
 Thus the anapole moment is zero for all the external particles on the
mass--shell. We have presented our results for the massless case alone
as it is sufficient to analyze the singularity structure of
SQED,~\cite{Ball-Chiu:1980}.
\\

\end{itemize}


The electromagnetic properties of elementary particles constitute
observables of important theoretical and phenomenological
interest, as they are model independent and respond to $P$, $T$,
and $C$ transformations. Therefore, they constitute windows through
which new physics effects could show up. In particular in neutrino
physics, they can act as a probe of whether neutrinos are Dirac of Majorana
particles. In this brief report, within the context of
scalar electrodynamics, we have shown that a charged scalar particle
develops an anapole moment at the one--loop level, which is finite
but gauge--dependent. Some kinematic limits have been
derived~\footnote{Incidentally, this form factor plays an important
role in constructing a non perturbative triple-gluon vertex
which would ensure gluon mass generation without seagull
divergences~\cite{Papavassilious:2009}. Although this form factor 
is logarithmically divergent as $q^2 \rightarrow 0$, it still diverges
sufficiently slowly for the argument constructed in~\cite{Papavassilious:2009}
to hold valid,~\cite{Bashir-Papavassilious:2009}.}.
In particular, it is shown that the anapole moment is zero for all
the external particles on the mass--shell. The existence of this
electromagnetic property is shown by using the effective Lagrangian
technique. We would like to comment that the anapole
moment would be the sole electromagnetic property of a scalar particle
that is neutral but differs from its own antiparticle, \textit{i.e.},
a particle which is associated with a non self--conjugate scalar
operator field.

\acknowledgments{ We acknowledge financial support from CIC under project no.
4.10 and CONACYT for the projects 46614-I, 50764, 94527
({\it Estancias de Consolidaci\'on}) and {\em Red de Cuerpos Acad\'emicos de
F\'isica de Altas Energ\'ias.}}


\begin{thebibliography}{99}
\bibitem{BCD} A. Bashir, Y. Concha--S\'anchez and R. Delbourgo,
Phys. Rev. \textbf{D76}, 065009 (2007).

\bibitem{Zeldovich} Ya. B. Zeldovich, Zh. Eksp. Teor. Fiz.
\textbf{33}, 1531 (1957) [Sov. Phys. JETP \textbf{6}, 1184
(1958)].

\bibitem{MN} J. Schechter and J. W. F. Valle, Phys. Rev. \textbf{D24},
1883 (1981); J. F. Nieves, Phys. Rev. \textbf{D28}, 1664 (1983);
L. F. Li and F. Wilczek, Phys. Rev. \textbf{D25}, 143 (1982); S.
P. Rosen, Phys. Rev. \textbf{D29}, 2535 (1984); S. M. Bilenky, N.
P. Nedelcheva and A. T. Petcov, Nucl. Phys. \textbf{B247}, 61
(1984).

\bibitem{ZB} A. Barroso, F. Boudjema, J. Cole and N. Dombey, Z.
Phys. \textbf{C28}, 149 (1985); F. Boudjema and N. Dombey, Z.
Phys. \textbf{C35}, 499 (1987); J. M. Hern\'andez, M.A Perez, G.
Tavares--Velasco and J. J. Toscano, Phys. Rev. \textbf{D60},
013004 (1999).

\bibitem{FB} F. Boudjema, C. Hamzaoui, V. Rahal and H. C. Ren,
Phys. Rev. Lett. \textbf{62}, 852 (1989); F. Boudjema and C.
Hamzaoui, Phys. Rev. \textbf{D43}, 3748 (1991).

\bibitem{Musolf} See for instance, M. J. Musolf and B. R.
Holstein, Phys. Rev. \textbf{D43}, 2956 (1991).

\bibitem{EL} W. Buchmuller and D. Wyler, Nucl. Phys. \textbf{B268}, 621
(1986). See also, J. Wudka, Int. J. Mod. Phys. \textbf{A9}, 2301
(1994).

\bibitem{ATLAS} G. Aad \textit{et al.} (ATLAS Collaboration), arXiv.0901.0512.

\bibitem{CMS} CMS Collaboration, CERN-LHCC 2006/001.

\bibitem{TTN} The one--loop electromagnetic properties of a
neutral vector boson have been studied in G. Tavares--Velasco and J. J.
Toscano, Phys. Rev. \textbf{D70}, 053006 (2004).

\bibitem{Ball-Chiu:1980} J.S. Ball and T-W. Chiu, Phys. Rev. \textbf{D22},
2542 (1980).

\bibitem{Papavassilious:2009} {\em ``Gluon Mass Generation Without Seagull Divergences"},
A.C. Aguilar and J. Papavassiliou, arXiv:0910.4142[hep-ph] (2009).

\bibitem{Bashir-Papavassilious:2009} Private communication of A. Bashir with J.
Papavassilious.

\end{thebibliography}
\end{document}